# Underwater Augmented Reality for improving the diving experience in submerged archaeological sites

Fabio Bruno[1,2], Loris Barbieri[1], Marino Mangeruga[1,2], Marco Cozza[2], Antonio Lagudi[2], Jan Čejka[3], Fotis Liarokapis[3], Dimitrios Skarlatos[4]

[1] Department of Mechanical, Energy and Management Engineering (DIMEG) - University of Calabria
Via P.Bucci, 87036 Arcavacata di Rende (CS) – Italy
e-mail addresses: (fabio.bruno, loris.barbieri)@unical.it

[2] 3DResearch S.r.l.
Via P.Bucci, 87036 Arcavacata di Rende (CS) – Italy
e-mail addresses: (marino.mangeruga, marco.cozza, antonio.lagudi)@3dresearch.it

[3] Human Computer Interaction Laboratory, Faculty of Informatics, Masaryk University
Botanická 68a, 602 00 Brno, Czech Republic
e-mail addresses: xcejka2@fi.muni.cz, liarokap@mail.muni.cz

[4] Photogrammetric Vision Laboratory, Department of Civil Engineering and Geomatics, Cyprus University
of Technology, 3036 Limassol, Cyprus;
e-mail address: dimitrios.skarlatos@cut.ac.cy

**Abstract**

The Mediterranean Sea has a vast maritime heritage which exploitation is made difficult because of the many limitations imposed by the submerged environment. Archaeological diving tours, in fact, suffer from the impossibility to provide underwater an exhaustive explanation of the submerged remains. Furthermore, low visibility conditions, due to water turbidity and biological colonization, sometimes make very confusing for tourists to find their way around in the underwater archaeological site.

To this end, the paper investigates the feasibility and potentials of the underwater Augmented Reality (UWAR) technologies developed in the iMARECulture project for improving the experience of the divers that visit the Underwater Archaeological Park of Baiae (Naples). In particular, the paper presents two UWAR technologies that adopt hybrid tracking techniques to perform an augmented visualization of the actual conditions and of a hypothetical 3D reconstruction of the archaeological remains as appeared in the past. The first one integrates a marker-based tracking with inertial sensors, while the second one adopts a markerless approach that integrates acoustic localization and visual-inertial odometry. The experimentations show that the proposed UWAR technologies could contribute to have a better comprehension of the underwater site and its archaeological remains.

# 1. Introduction

The Mediterranean Sea has a huge cultural and archeological asset, consisting of ancient shipwrecks and sunken cities, with broad potential for the development of the tourism sector. Furthermore, the latest advances in the field of survey techniques for the exploration of the seabed is exponentially increasing the discovery of underwater cultural heritage (UCH) sites. Nevertheless, many of them are not accessible because of the limitations due to the environmental context, such as depth of the site and sea currents, or to local and international laws and regulations. Furthermore, those that can be visited by diver tourists present some issues related to the marine environmental conditions that do not permit a satisfactory exploitation of the underwater archaeological sites. Simultaneously, the UCH has provoked considerable interest thanks to the work carried out in the recent years by the National Commissions for UNESCO that discourages the adoption of the traditional excavation and recovery methods in favor of on-site examination and in situ preservation and conservation techniques [UNESCO 2001].

To this end, computer graphics techniques like 3D reconstructions, Virtual Reality (VR) and Augmented Reality (AR), have demonstrated to be a highly effective means of communication for facilitating the access and increasing the value and the public awareness about the cultural heritage. In fact, in the last decades, a number of researchers are testing and perfecting reconstruction techniques and developing new technologies for the exploitation of the UCH [Chapman et al. 2008; Haydar et al. 2011; Bruno et al. 2018, 2016a, 2016b]. Thanks to the advances achieved in the field of photogrammetric reconstruction techniques, it is now possible to make high-resolution and accurate 3D reconstruction of the underwater scene with low-cost technology and in a relatively short time [Skarlatos et al. 2012; Cebrián-Robles et al. 2016; Lagudi et al. 2016; Łuczyński 2017].

Despite these achievements and significant progress made in the last years, VR- and AR-based applications for improving the diving experience in the underwater archaeological sites are still few, with many shortcomings to overcome and huge development potentials to unlock.

A good help for understanding the real extent of these potentials might be the fact that, due to water turbidity and biological colonization, in the submerged archaeological sites the divers often suffer from low visibility conditions and this leads to a less understanding of the underwater environment and a higher probability for them to miss the sense of direction. Unfortunately, GNSS sensors (GPS, GLONASS, and Galileo) are inadequate to this end since their signals are absorbed in water after a few centimeters below the sea level. Furthermore, guided or accompanied archaeological diving tours are carried out with experienced divers, but it is not possible to perform a fluid and direct communication unless they use full-face diving masks or analogous dedicated equipment. At the moment, there are few attempts to support the divers by facilitating their comprehension of the archaeological context. One of these has been implemented in the underwater archaeological site of Punta Scifo, located in the East coast of Calabria at 10 km far from Crotone, where an underwater trail (Fig.1a), consisting of a guide rope and floating labels fixed at the margins of the archaeological remains, permits divers to know their position, identify artifacts and read the correspondent information printed on plastic slates (Fig.1b).

Another example can be found in Sicily where the Superintendence of the Sea has implemented, in seven underwater archaeological sites, interactive itineraries by identifying the archaeological remains through a small float with a Quick Response (QR) code label that allows divers to get access to the historical and archaeological information employing a handheld waterproof QR code scanner. There are then some fruitful examples, but they are fairly simple and still do not exploit the existing potentials.

On the basis of the abovementioned considerations, AR technologies could be a useful tool to overcome these limitations and could be a valid solution to improve the readability and understandability of the submerged archaeological sites and enhance the overall diving experience by providing interesting information about the ancient remains and artifacts.

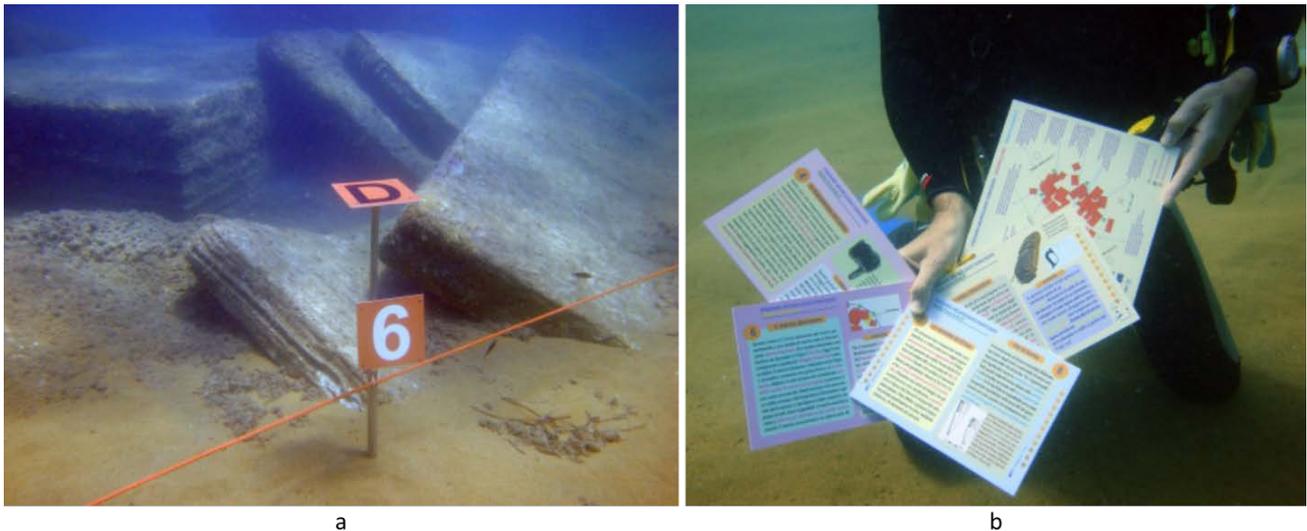

**Figure 1.** Underwater itinerary in the submerged archaeological site of Punta Scifo. (Images courtesy of the Marine Protected Area of "Capo Rizzuto")

In this regard, the paper investigates the feasibility and potentials offered by the AR technologies for improving the diving experience in the underwater archaeological sites, and provides an overview of the first result achieved in the iMARECulture project. In particular, the Horizon 2020 iMARECulture (Advanced VR, iMmersive Serious Games and Augmented REality as Tools to Raise Awareness and Access to European Underwater CULTURal hEritage) project (www.imareulture.eu) [Skarlatos et al. 2016; Bruno et al. 2018] aims to investigate and develop AR-based solutions for promoting and improving the public awareness about the UCH. This aim will be reached according to two different technologies that adopt hybrid tracking techniques to perform an augmented visualization representing the actual conditions of the ancient ruins in the underwater site and a hypothetical 3D reconstruction of the archaeological remains as they appeared in the past of the Roman era by means of a commercial tablet housed in a waterproof case. The first technology presented in the paper integrates a marker-based tracking with inertial sensors, while the second one adopts a markerless approach that integrates acoustic localization and visual-inertial odometry.

The paper is organized as follows. In Section 2 the state of the art is presented. The case-study adopted for the preliminary field tests is described in Section 3. Section 4 and Section 5 details the AR technologies developed in the iMARECulture project to improve the divers' experience in the submerged archaeological sites. Finally, Section 6 gives the conclusions of the paper.

## 2. State of the art

AR technology has demonstrated to be a very useful tool for improving the visitor experience in cultural sites since it provides visual information contextualized with the environment and with the user point of view. Through the use of marker or location-based AR applications, tourists can orient themselves inside large areas, receive multimedia contents seamlessly, understand better the cultural value of what they are observing, visualize hypothetical reconstruction of monuments and objects to represent them as these appeared in the past. All these opportunities have been well exploited for the terrestrial cultural heritage while, up to now, the use of AR still remains completely unexplored in the context of the UCH.

In this context, in the last years, various researches are investigating and proposing different frameworks for the reconstruction, collection, and visualization of the UCH but for its exploitation outside of the underwater environment [Chapman et al. 2008; Haydar et al. 2011; Varinlioğlu 2011; Gallo et al. 2012; Katsouri et al. 2015].

About Underwater Augmented Reality (UWAR) the first type of application was developed for military purpose in 1999 [Gallagher 1999], it consists of an underwater head-mounted display (HMD) for Navy divers that allows to augment the diver's view with virtual information in military operations, especially under poor visibility conditions. However, this is not an AR application in a strict sense, since the virtual information presented is not registered with the user's 3D perception of the real world.

A more sophisticated system, developed for edutainment purposes, was presented in 2009 [Morales et al. 2009], it consists of a UWAR system, based on optical square-markers, that provides visual aids to increase divers' capability to detect, perceive, and understand elements in underwater environments. A similar wearable waterproof system, but limited to a swimming pool environment, was developed in the same year by the Fraunhofer Institute for Applied Information Technology [Blum et al. 2009; Oppermann et al. 2016]. Another marker-based AR underwater device, that can be adopted in swimming pools, for aquatic leisure activities is the Dolphyn system [Bellarbi et al. 2013]. This system provides AR contents to the user through a tablet, housed in a waterproof case, that has been equipped with GPS and wireless systems. Body movement is restricted because it is necessary to interact with the equipment with both hands.

When compared to visibility conditions in swimming pools, the precision of computer vision algorithms is impaired by bad visibility conditions in the sea. An impact of such conditions on a set of open-source marker detection algorithms was measured in laboratory conditions [Cesar et al. 2015], but fortunately, their performances can increase by using offline image enhancing algorithms [Žuzi et al. 2018] and real-time algorithms [Čejka et al. 2018, Čejka et al. 2019]. Registration of objects in images can also be improved for the purpose of underwater photogrammetry [Sarakinou et al. 2016, Agrafiotis et al. 2017]. These works, however, focus only on specific parts of AR, namely on the detection and recognition of objects. To our knowledge, there is no work that would evaluate a complex underwater AR system in conditions of an open sea.

These efforts demonstrate a strong interest in the research community for the development of the UWAR since this technology could be applied to a large variety of sectors that operate in the marine environment. Nevertheless, as above mentioned, the progress made until now in this area is insufficient and their main limitation is due to the tracking capabilities of the systems adopted for the underwater environment. Indeed, algorithms that face the problem of underwater tracking have mainly been investigated in relation to underwater vehicles and robots [Tan et al. 2011; Costanzi et al., 2016]. Most of these solutions, based on a dead reckoning approach [Allotta et al., 2016], are intended for the use in wide marine environments and are therefore focused on an approximate large distance tracking. Furthermore, these solutions are not able to accurately estimate the user's 6DOF pose and therefore to be used for accomplishing the correct alignment between virtual objects and real underwater world.

## 3. The case-study: the underwater archaeological park of Baiae

The UWAR technologies developed in the iMARECulture project has been intended and tested for the Marine Protected Area - Underwater Archaeological Park of Baiae, located in the volcanic area of the Phlegrean Fields, a few kilometers North of Naples (Italy). This is a worldwide known site because it is a typical representative of the phenomenon of bradyseism as the rests of the Roman age are actually at a depth ranged from 0.0 m to 15.00 m from the sea level, and only a few ruins are still on the coastline, inland. The Underwater Park of Baiae (Fig.2) is famous also for its extensive submerged area of 176.600 hectares, and the wide range of different architectural structures, i.e., fisheries and harbor buildings, thermal baths, residential buildings, and villas, with some decorations that are still preserved. In particular, the experimentation has been undertaken in the complex of the "Villa con ingresso a protiro - Villa with Vestibule" dated to the first half of the II century AD.

The use scenario to which the UWAR technology developed in the iMARECulture project is intended consists in providing to the divers the possibility to know their position within the submerged environment and to enable the augmented visualization of the actual conditions of the ancient ruins in the underwater site and a hypothetical reconstruction of the villa, thus easily understanding the luxury and the importance of that building during the Roman era.

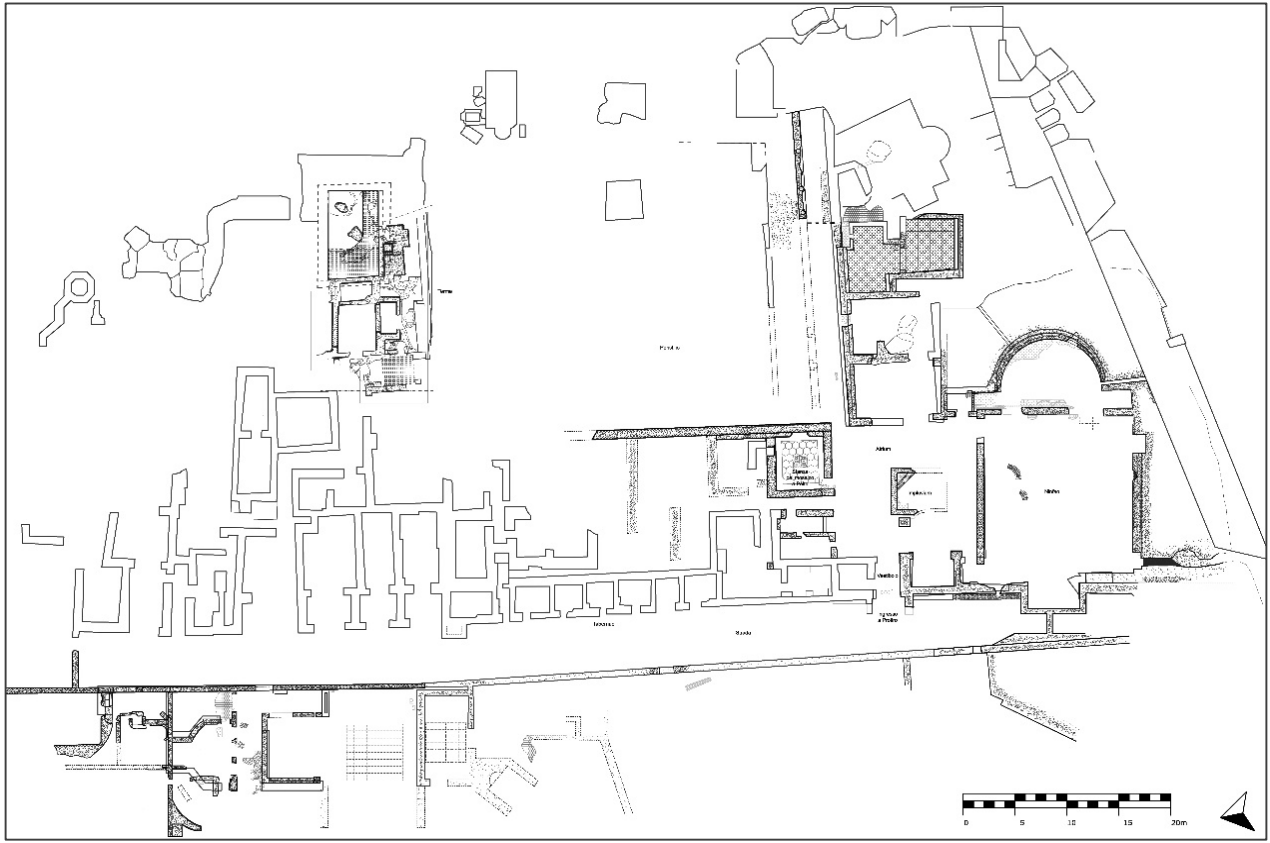

**Figure 2.** Planimetric archaeological map of "Villa con ingresso a protiro". (Image courtesy of ISCR)

The starting point for preparing the 3D data and contents that are necessary for the functioning of the proposed UWAR technologies consists of the 3D reconstruction of the abovementioned archaeological area that has been carried out by combining optical and acoustic techniques [Lagudi et al. 2016; Mangeruga et al. 2018]. The 3D reconstruction model is then populated with a number of points of interest (POIs), placed on the seafloor, which provide the position of the distinctive and characteristic element of the specific underwater site. In particular, the POIs are represented with different colors depending on the category they belong to, e.g., yellow for the historical and archaeological artifacts and remains and green for biological organisms. Thanks to this data the underwater tablet provides to the user: a map of the underwater scene that allows the diver to know his/her position within the submerged site; archaeological, historical and biological information about the specific archaeological context; and an enhanced diving experience through an on-site augmented visualization representing the actual conditions of the ancient ruins in the underwater site and of a 3D hypothetical reconstruction of the "Villa con ingresso a protiro".

About the 3D hypothetical reconstruction, this has been achieved by means of a theoretical and multidisciplinary scientific approach [Davidde Petriaggi et al., 2018], under the direction of Barbara Davidde Petriaggi, that exploits the high-resolution 3D data together with drawings and other historical and archaeological information to build a suggestive and consistent digital reconstruction of the underwater architectures not anymore existing. In particular, the reconstruction process starts with gathering historical documentation, scientific literature and geometric data (archaeological maps, illustrations, photos, Digital Terrain Model, etc.). All the data are then analyzed and put in relation by the experts to generate and investigate different interpretation hypotheses that are validated by means of an iterative critical revision. The process, in fact, is based on interleaving a phase of technical reconstruction with a strong critical revision in order to generate a feedback process, iterating the construction /correction loop as much as needed. Finally, to map the evolution of the virtual interpretation, several 3D layers are saved together with the final model, examined and approved by the scientific experts. The following figure depicts the final 3D hypothetical reconstruction, as it appeared in the past, of two different rooms of the complex of "Villa con ingresso a protiro", and in particular of the *atrium* with *impluvium* (Fig.3a) and the room with pelte mosaic (Fig.3b).

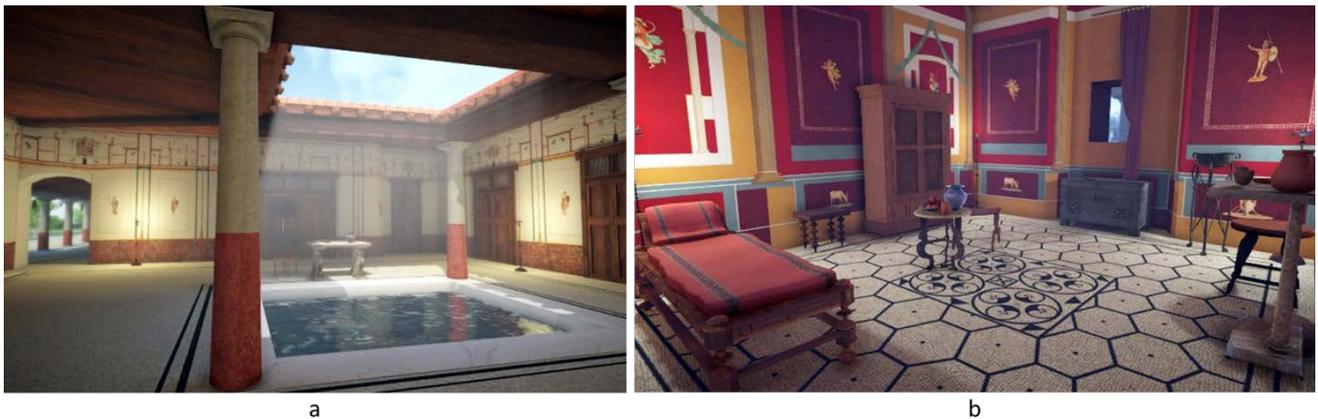

**Figure 3.** Different portions of the 3D hypothetical reconstruction of the Villa: atrium with impluvium (a), room with pelte mosaic (b).

## 4. Marker-based underwater AR

AR-based applications require a precise location of the viewer in real-time to properly superimpose virtual objects into the real scene. One of the possible solutions is to use markers, which are artificial objects that are easy to detect [Čejka et al. 2018]. These solutions are fast, cheap and reliable, however, their main downside is that the area must be calibrated before each dive (e.g. populated by such markers).

A hybrid tracking approach is one of the tracking solutions used in the iMARECulture project and presented here. In particular, the solution builds on a marker-based tracking that uses visual information from the front camera of a smartphone or tablet, and combines it with the data from the inertial measurement unit (IMU) of the device, namely with the data from the accelerometer and gyroscope.

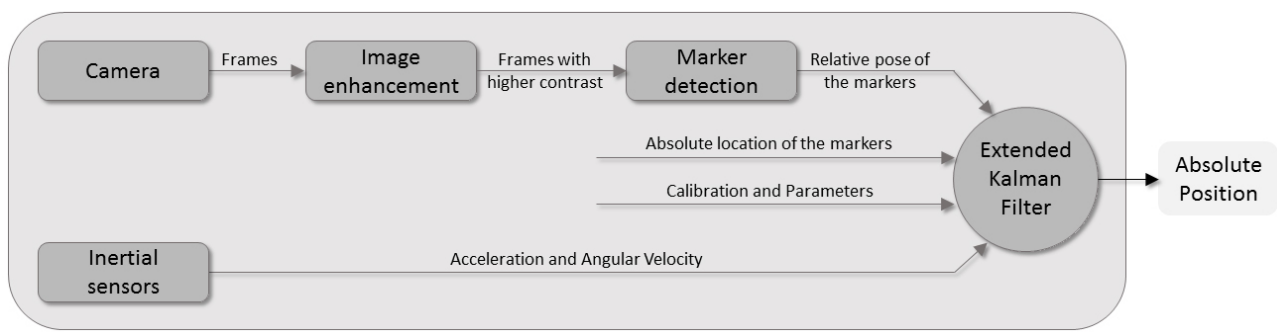

**Figure 4.** Architecture of the marker-based tracking system

An overview of the system architecture is presented in Figure 4. The front camera of the device, which is sealed in a waterproof housing, records images of surrounding objects, and improves them in real-time to increase their contrast. The system also allowed the diver to change between methods that improve images, namely CLAHE [Pizer et al. 1987], debluring [Krasula et al. 2017], white balancing [Limare et al. 2011], and white balancing that is adapted to underwater sea environment [Čejka et al. 2018]. Then, the AR system detects squared planar markers in these images using the standard computer vision libraries ArUco [Garrido-Jurado et al. 2014], which is available as a part of OpenCV library, and computes the relative position of the user (which is in essence the camera of the device) to each marker. The application calculates the absolute position of the diver when the markers are placed at pre-defined locations (e.g., center of the mosaic of Villa con ingresso a protiro). This visual information is combined with data obtained from the inertial sensors of the device using an Error-state Extended Kalman filter [Solà 2017, Neunert et al. 2016]. This filter reduces the noise present in visual and IMU data, and ensures that the computed movement is smooth. It is also able to estimate the position of the device when the markers are lost from the line-of-sight. Having no line-of-sight with the markers is typically happening due to bad visibility conditions in the sea, but also due to fast movements of the camera and moving the markers out of its field of view. Calibration of the camera and sensors is performed before the system is deployed on the site. Distortion parameters of the camera depend on the environment, and are obtained underwater using a calibration board and computed offline. Noise characteristics of the marker detecting algorithm and inertial sensors, required for the Kalman filter, are the same on land and underwater, so they are measured in laboratory conditions in a stable and controlled environment by computing statistical parameters of the results when the application is left running several minutes with no movements of the device.

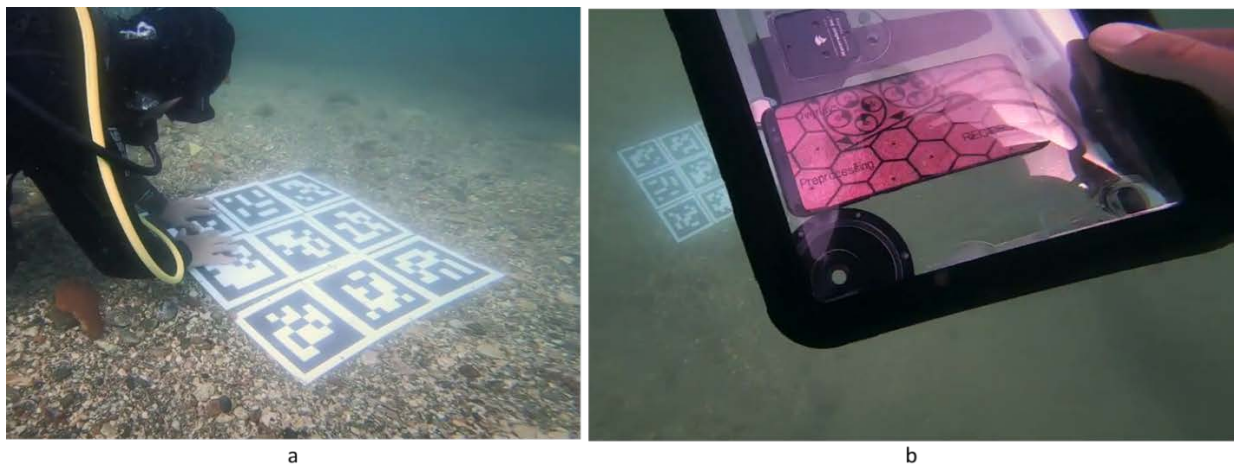

**Figure 5.** A grid of 3x3 markers is placed at the location of Villa con ingresso a protiro (a); AR model of the villa rendered at the location of the markers (b)

An experimental evaluation of the AR system was performed at the location of Villa con ingresso a protiro. For this test, 9 optical square markers with a size of 19 x 19 cm were placed into a 3 x 3 grid (Fig.5a). The diver was equipped with a Samsung S8 device housed in a waterproof case, which recorded the data from the sensors at frequency 60 Hz and the data from the camera at frequency 30 Hz in resolution 1280 x 720. The Samsung S8 device has an octa-core CPU type with a speed of respectively $4\times2.3$ GHz and $4\times1.7$ GHz which is enough to get an update rate of 20 Hz, but usually 30 Hz. They swam around to examine the virtual model of the Villa underwater. When the system detected the markers, it placed the virtual 3D reconstruction of the villa at their location (Fig.5b). The application also provides two methods of visualizing the 3D reconstruction in AR, a solid textured representation, and a wireframe representation.

During tests, the data from the camera and inertial sensors were recorded and stored uncompressed into a memory of the mobile device. Such data were used for further offline evaluation to compare multiple marker detectors and image improving algorithms to find an optimal solution [Žuži et al. 2018, Čejka et al. 2018, Čejka et al. 2019]. The comparison showed that real-time improving of images is sufficient to increase the distance at which the markers are visible, which was also confirmed by the divers. Offline comparison also showed that Extended Kalman filter smooths the computed movement of the diver, and eliminates abrupt changes in position caused by errors in the detection of markers (Fig.6). Measured trajectory of the divers also helps to understand their behavior on place.

A similar experiment was also performed in a laboratory to measure the absolute error in position and orientation. The ground truth trajectory was obtained by OptiTrack motion capture system, which recorded the position of the mobile device at frequency of 120 Hz and precision under 1 millimeter. Without using Kalman filter, the average position error per frame was 52 mm and average orientation error was 1.9 degrees. With Kalman filter, average position error decreased to 46 mm and average orientation error decreased to 1.0 degrees. Increased precision in orientation is very important as this error is more apparent at distant objects like walls of the virtual room – an error of 1 degree in orientation corresponds to approximately 35 mm error in position of an object that is 2 meters away.

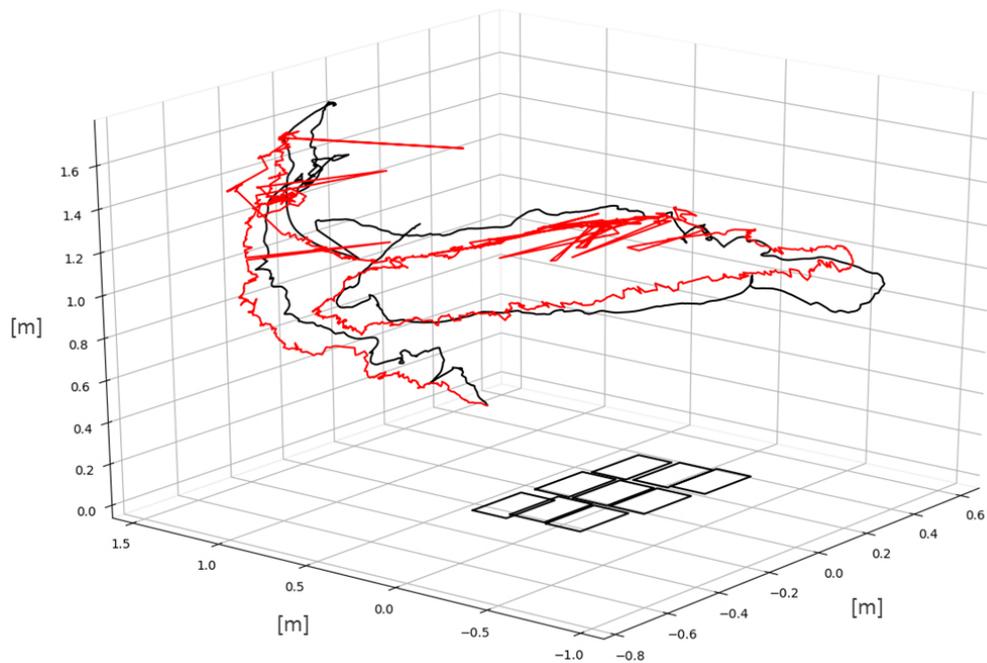

**Figure 6.** Trajectory of the diver during the experiment, obtained solely by marker detecting algorithm (red), and using a fusion of the visual data from the camera and inertial data from sensors with the Extended Kalman filter (black). Notice that the Extended Kalman filter smooths the trajectory and avoids spikes in movement caused by errors in the detection of markers.

## 5. Markerless underwater AR

In addition to the marker-based technology, the iMARECulture project has led to the development of a markerless UWAR technology which augmented visualization is provided to the diver by means of a commercial tablet, housed in a waterproof case (Section 5.1). Unlike the marker-based approach, the developed markerless UWAR, on the one side, does not require to prepare the scene by placing fixed visual markers on the seabed to estimate the orientation and position of a camera with respect to the real world frame, and furthermore it eliminates the need to detect markers in order to perform the superimposition of the virtual elements. But, on the other side, the quality of the augmented visualization provided through the markerless approach is strictly related to the precision of the diver's position tracking, and consequently to the underwater acoustic positioning systems which suffer from low update rate and low accuracy.

Since AR visualization requires a high frame rate to operate properly, it is quite evident that such acoustic localization systems alone are inadequate for this purpose. Then, in order to overcome this limitation, a hybrid tracking system (Section 5.2) has been specifically developed by integrating acoustic localization and visual inertial odometry to enable a consistently high frame rate and provide to the user a consistent and smooth AR visualization.

## 5.1. Underwater tablet

The hardware developed to make markerless AR accessible underwater (Fig.7a) consists of a commercial tablet, housed in a waterproof case and coupled with an acoustic localization system, implemented by AppliCon Srl. The waterproof case allows the diver to fully interact with the tablet's touchscreen through his/her fingers thanks to a special flexible membrane. In particular, all the touchscreen functionalities are preserved thanks to a pressure management system that ensures the presence of an air gap between the tablet display and the housing membrane. The tablet adopted is a commercial iOS-based device, namely a 9.7-inch iPad. Moreover, additional electronic devices, that include the transponder of the acoustic localization system, are housed in a cylindrical waterproof case (Fig.7a) fitted on the back side of the tablet housing. All these devices communicate between themselves through a WiFi network. Since WiFi communication works underwater within a range of few centimeters, this solution has been adopted for the communication between the acoustic transponder and the tablet in order to avoid the use of wiring that could have caused water tightness issues.

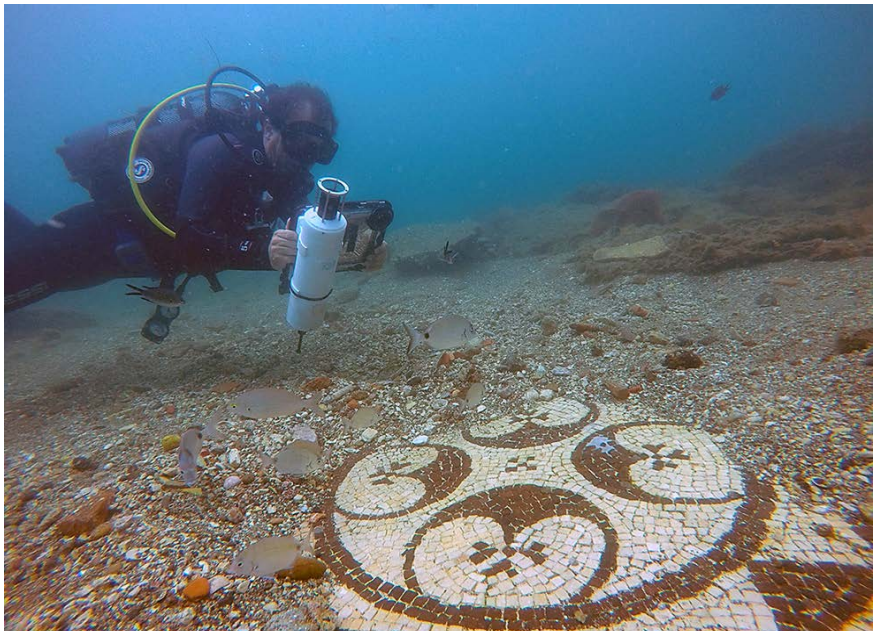

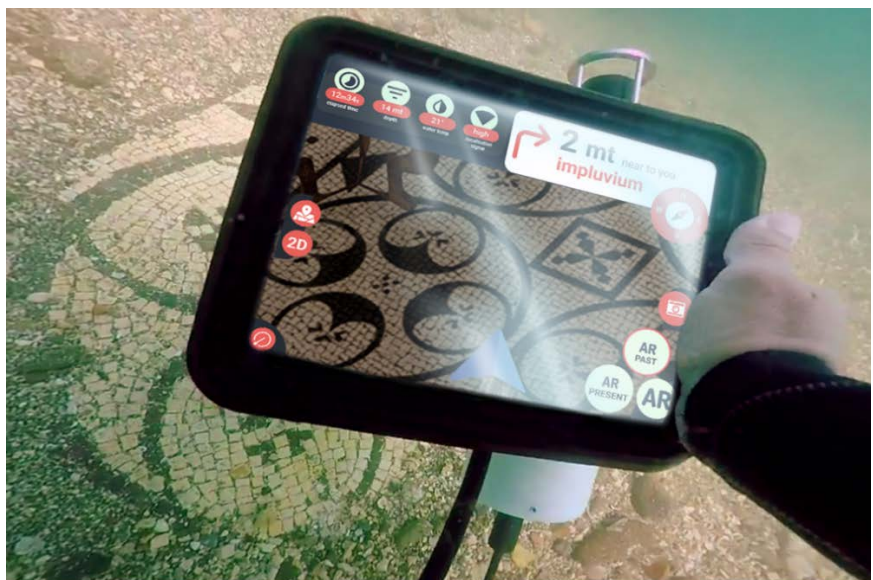

**Figure 7.** The underwater tablet during the test in the "Villa con ingresso a protiro".

The UWAR is provided to the user through the underwater tablet by means of an application developed in Unity 3D [Unity]. This is a real-time cross-platform engine developed by Unity Technologies that provides workspace and complete toolset for the creation of interactive 3D content. The tracking functionalities and AR capabilities have been implemented by means of the ARKit [ARkit] extension for Unity. In particular, the Unity framework accesses to the Operating System's features to perform several tasks, such as the access to the device's camera or sensors. The core of the UWAR application contains all the modules that handle its main functionalities of the UWAR app, such as 3D map management, AR interactions, geolocation, communication with the modem, cloud access over the internet, and user interfaces (UIs) for user interactions.

All the functionalities are placed in specific areas of the screen in order to structure, label and organize the content so that users can find exactly what they need to perform the task they want and to reach their goal. Assuming that the diver interacts with the device while keeping the housing with one hand, all the interaction buttons are placed on the left and right side of the screen. The different areas of the UI allow the user to: obtain information about the current dive session; obtain information about the POIs; control the view of the map of the underwater site (between the third-person-view and the top-view); access to the camera features; enable the AR visualization.

In particular, as depicted in figure 7b, the diver's position is shown by the means of a 3D arrow while the On-Board Computer area in the top-left corner shows the elapsed time, the current depth from the sea surface, the water temperature, and the acoustic signal strength. In the top-right corner, a compass shows the orientation of the tablet with respect to the North. On the left side of the screen, a menu allows the user to control the zoom level of the 2D map of the underwater site. The camera button, placed on the right side of the display, permits to control the tablet's embedded camera. The command button "AR" on the bottom-right of the screen allows the user to enable the augmented visualization. In particular, when the "AR" button is selected, other two button appear on the screen to allow the user to switch between the augmented visualization representing the actual conditions of the ancient ruins in the underwater site (Fig.8a) and the hypothetical 3D reconstruction of how the site appeared in the past (Fig.8b) during the Roman era.

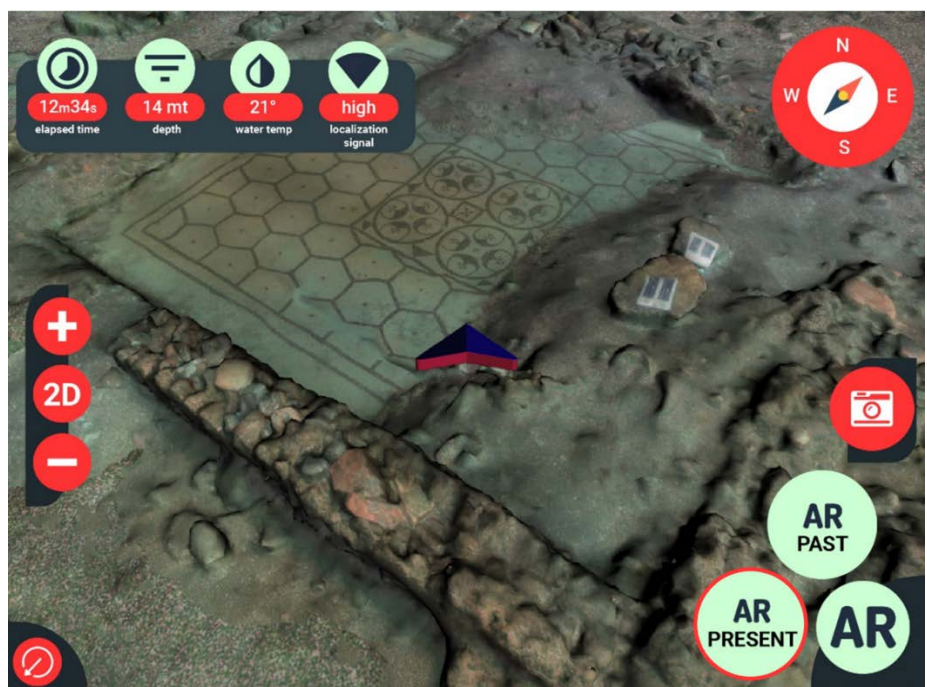

a

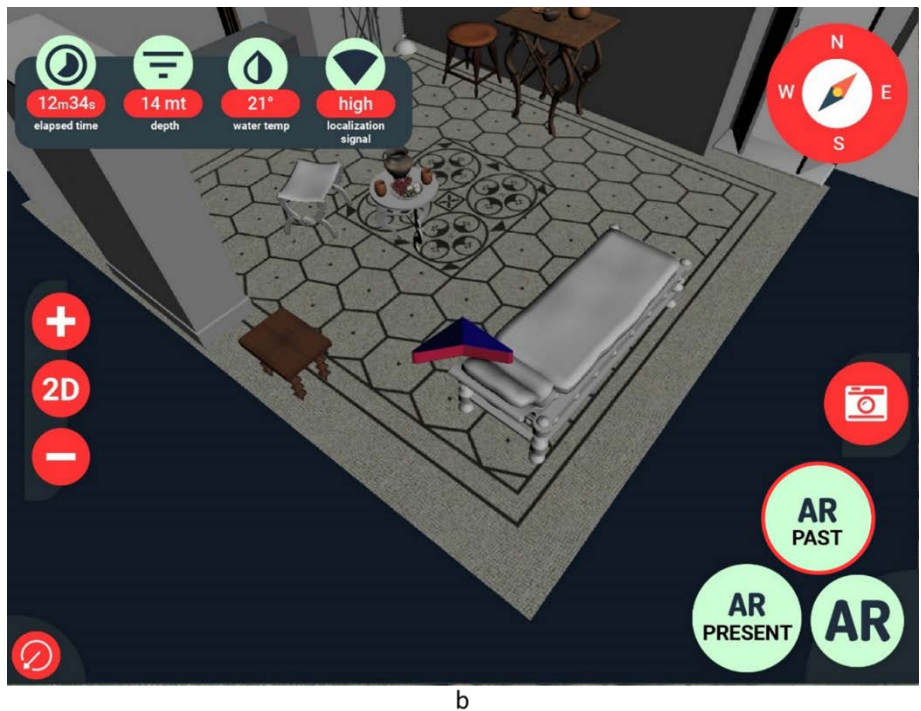

**Figure 8.** UI of the UWAR App showing the augmented visualization of the actual condition (a) and the hypothetical reconstruction of the underwater site (b).

It is worth noting that, as depicted in figure 8, the augmented virtual models are not superimposed on the frames captured by the tablet camera as would be expected in a classical AR. This choice was due to the low picture quality obtainable in the majority of the underwater sites such as the Underwater Archaeological Park of Baiae. In fact, the imagery produced in this kind of environment suffers from a lack of contrast and poor visibility due to the particles suspended in the water.

### 5.2. Markerless hybrid tracking

As abovementioned, the update rate of the acoustic positioning system alone is around 0.2 Hz, and this is too low to deliver a seamless AR experience due to the long delay between two subsequent positions provided by this system. Furthermore, whatever the acoustic positioning system suffers from packet loss that further compromises the rate and, therefore, the quality of the acoustic positioning. The acoustic communication is also degraded by the presence of the diver that during the immersion can frequently place itself between the acoustic transducer integrated with the tablet and the one on the sea surface. This increments the packet loss issue and decreases the update rate so that even an acoustic positioning system with a formal higher update rate cannot overcome the delay between two subsequent positions. In order to overcome this limitation and improve the performance of the proposed UWAR technology, the acoustic positioning system has been integrated into a hybrid tracking system which merges positioning data, generated by the acoustic system, with data coming from a Visual Inertial Odometry (VIO) framework. In particular, given the low update rate of the acoustic system, it has been implemented a data fusion strategy aimed to fill the gaps between two consecutive acoustic positioning data.

The developed markerless hybrid tracking system's architecture is shown in the following figure (Fig.9). As abovementioned, it is composed of two main sub-components: an acoustic positioning system and a VIO framework that is meant to bridge the gap between two consecutive acoustic positions.

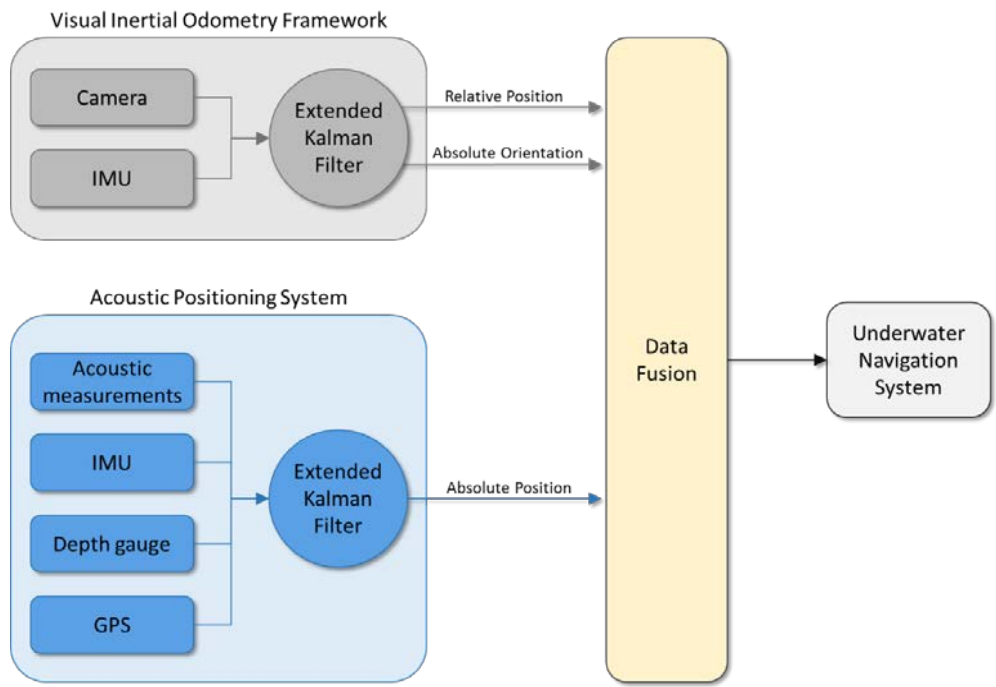

**Figure 9.** Markerless hybrid tracking system's architecture.

As depicted in figure 9, the acoustic positioning system does not rely only on acoustic measurements, but also on a depth gauge, GPS and IMU sensors in order to compute the absolute position of the diver. Obviously, due to the limitations of GPS, this sensor is placed outside of the underwater environment. At the same time, the VIO framework combines data from the camera and the inertial platform of the tablet to calculate the absolute orientation and relative position with respect to the starting point. In particular, the VIO framework recognizes notable features in the scene image, tracks differences in the positions of those features across video frames, and compares that information with motion sensing data. The result is a model of the device's location and motion. As regards the VIO framework, it has been decided to employ ARKit by Apple that is based on the SLAM (simultaneous localization and mapping) technology that allows to recognize the scene from individual points.

About the fusion of the different typology of data, it has been employed a strategy that consists of using the acoustic positioning data as an initial reference point for the VIO framework that, as previously stated, delivers only local positioning data with respect to a starting point. Whenever a new acoustic positioning data is available, the VIO framework is reset and its reference point is moved to the new acoustic positioning data, while the hybrid tracking system relies only on this acoustic data for the localisation. Until no new acoustic positioning data is available, the hybrid tracking relies on the VIO in order to calculate the actual position. By fusing the data provided by these two positioning systems through the strategy previously described, it's possible to fill the gap between two consecutive acoustic positions. This data fusion strategy presented is only a preliminary approach that will be enhanced in the future. For instance, a smarter data fusion strategy could be employed by means of a Kalman filter that could predict in a better way the position of the diver given all the positioning data from acoustic and VIO systems.

The following figure (Fig.10) shows in detail as the different positioning data are fused in the hybrid system. The absolute orientation is provided exclusively by the ARKit framework, while the "Z" coordinate, i.e., the depth, is computed by the means of only the pressure sensor integrated with the acoustic system. No acoustic signal is employed to acquire information about the current depth. The "X" and "Y" coordinates are computed by the data fusion of the two positioning systems.

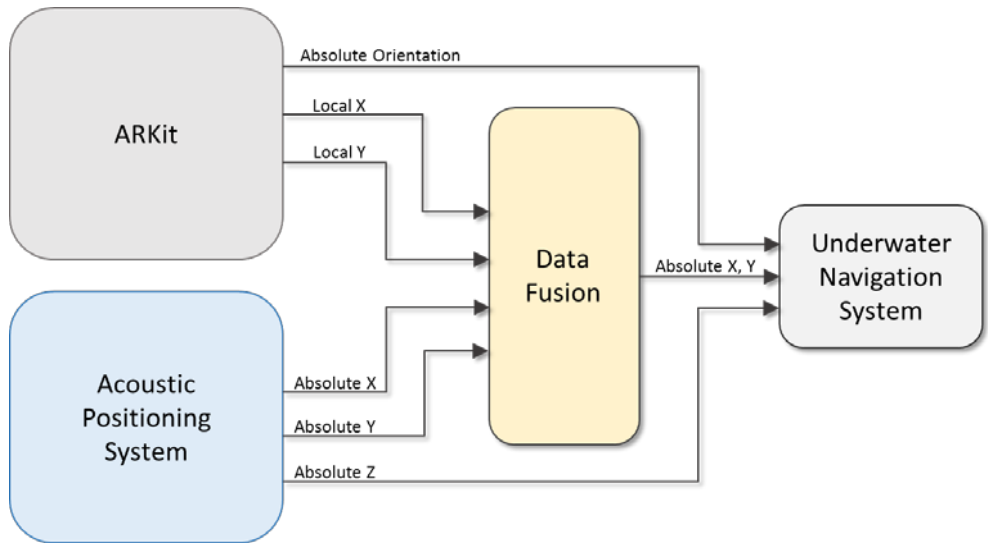

**Figure 10.** Deepening on the architecture of the hybrid tracking system.

### 5.2.1. Preliminary field test

A preliminary field test of the hybrid tracking system, aimed to estimate its performances, has been carried out in the shallow water of the underwater archaeological site of Baiae. The ARKit framework was employed for the visual tracking, while the SeaTrac USBL, manufactured by Blueprint Subsea, was used for the acoustic tracking. The manufacturer declares that this USBL system has an acoustic range of 1km and a range resolution of ±50mm. No information about the update-rate and the accuracy of this acoustic system has been released.

Figure 11 shows the UI's app that has been modified to meet the needs of the test. In particular, some buttons have been removed, and some panels have been introduced in order to view the real-time camera images, the ARKit status, and some debugging values provided by the acoustic positioning system.

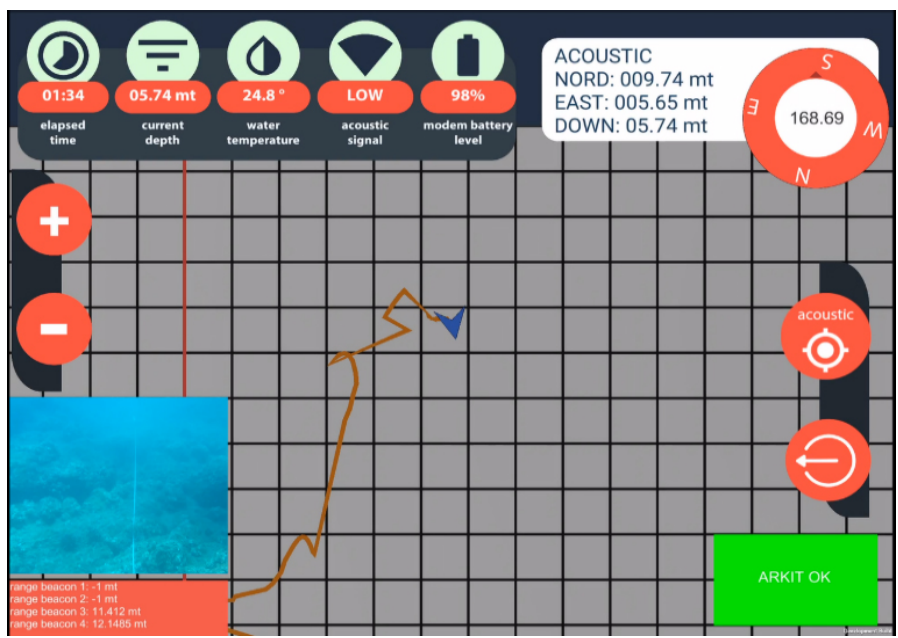

**Figure 11.** Modified UI of the UW tablet for field test experimentations.

This underwater test has been designed to evaluate the capability of the hybrid tracking system to bridge the gap between the calculation of two consecutive acoustic positions. Moving around the underwater site, the estimated position of the framework was compared to a ground-truth (pre-defined know points in a path). The protocol developed for the execution of the preliminary test consists of the following six steps:

1. construction of the underwater ground-truth. It was necessary to compare the collected positioning data with a known path in order to evaluate the performance of the hybrid tracking system. The path is square-shaped (Fig.12a), with edges marked with labeled panels (A, B, C, and D), and sides traced with graduate ropes.
2. Calibration of the acoustic localization system carried-out on the boat.
3. Deployment of the acoustic positioning system in the sea. Usually, the USBL acoustic positioning method involves measuring the range from a vessel, on which the USBL's transceiver is placed, to a single subsea transponder. Then, the overall positioning error is affected by the positioning error of the ship's GPS and the positioning error of the transponder relative to the ship's position. To cancel the effect of the GPS's error, the local beacon of the USBL system has been placed on the vertex A (Fig.12a) at an approximate distance of 3 meters above the seabed.
4. Measurement of the depth of the local beacon.
5. Annotation of the actual size of the square and its deviations respect to the planned ground-truth. This step is important for having a precise reference for the comparison and evaluation of the data measured during the test.
6. Run the test by starting the app when the diver is on the vertex A. Then the diver performs a complete counterclockwise lap of the squared ground-truth, passing through the B, C, D vertices, and ending on the starting vertex. The test is carried at a constant depth of about 6 meters deep below sea level.

The app has been implemented to automatically save the acoustic, optical and sensor data, measured during the test, into a log file for their next analysis and evaluation.

The real ground-truth built in the underwater site was slightly different from a square-shaped path because of the difficulties encountered in the underwater environment. The actual dimensions reported at the end of its deployment were: $|AB| = 30m, |CD| = 30m, |AD| = 29.26m, |BD| = 43.3m,$ and $|AC| = 41m$. Moreover, the AC side was tilted about 11 degrees with respect to the north direction.

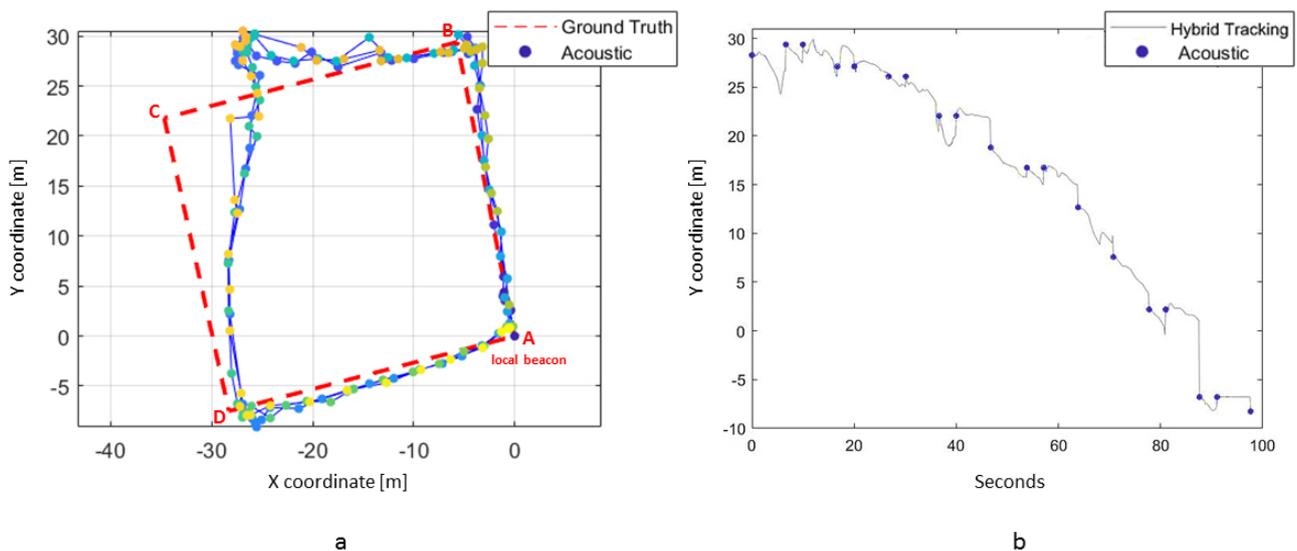

**Figure 12.** Acoustic data acquired moving along the ground-truth (a), and comparison of acoustic and data tracked according to markerless hybrid approach (b).

Figure 12 shows the result of the test. In particular, figure 12a depicts the empirical acoustic data acquired during three consecutive counterclockwise laps of the planned ground-truth. The ground-truth is represented with a dotted red line, while the acoustic data are represented in time progression through a scatter colour that goes from violet (start) to yellow (end of the experiment). The blue lines represent the connections between two consecutive acoustic data. The results show that on the upper left edge (edge C in figure 12a) of the squared ground-truth the error of the acoustic data exceeds 5 meters and this is too much for performing an acceptable augmented reality visualization. Since this error appears in all the three laps it is probably due to multipath propagation which becomes even more evident in the case of shallow waters. As a consequence, this preliminary test has pointed out that USBL systems are not adequate to perform the proposed application of UWAR in shallow waters. In fact, as shown in figure 12a, the USBL works quite well along the planned ground-truth except for some points that make the acoustic positioning system unstable. For this reason, in the next experimentations also SBL (short baseline) and LBL (long baseline) acoustic positioning systems will be taken into consideration in order to verify if these solutions allow to improve the performance of the acoustic tracking when the seabed is a few meters deep. Nevertheless, an interesting outcome of the hybrid tracking test is depicted in figure 12b in which the improvement of the tracking becomes evident thanks to the adoption of the VIO. In particular, this figure plots the distance-time graph of only the "Y" coordinate of the CD segment (Fig.12a) in order to better highlight the impact of the VIO on the tracking of the diver' position. In particular, the graph shows the acoustic positioning data, acquired by means of the SeaTrac that has a low update rate of approximately 0.2 Hz, and the hybrid tracking data that, thanks to the VIO framework which has an update rate up to 60 Hz, provides a graphical continuity to the path. This continuity is good in some cases (from 28 to 34 seconds, 47-55, 70-77 and 80-88) and not always satisfactory in others (from 37 to 40 seconds, 55-58 and 77-80). However, this preliminary results are encouraging and demonstrate that it is then possible to perform a consistent and smooth UWAR visualization by increasing the update rate from the 0.2 Hz of the acoustic positioning system alone up to the 60 Hz of the hybrid tracking system.

## 6. Conclusions

The paper has presented two novel UWAR technologies that can improve the divers' experience in submerged archaeological sites. In particular, the proposed technologies, developed in the H-2020 funded iMARECulture project, provide to divers, through the interaction with a tablet-based system, their position over the 3D map of the underwater archaeological site, an augmented visualization representing the actual conditions of the ancient ruins in the underwater site and another augmented visualization representing a hypothetical 3D reconstruction of the archaeological remains as they appeared in the past during the Roman era.

The paper has presented also the results of the preliminary field test carried out in the Underwater Archaeological Park of Baiae to assess the feasibility and practical potentials of the proof of concept of the developed UWAR technologies. In particular, the field test made it possible to confirm the proper functioning of the adopted visual tracking techniques in the underwater environment notwithstanding the negative effects of the water turbidity and refraction that occurs at the air-glass-water boundary. The test produced encouraging results that have been achieved both for the marker-based approach, with the adoption of the ArUco libraries, and the markerless approach that implements the latest ARKit libraries based on the SLAM technology to recognize the scene from individual points.

It is worth noticing that since the overall positioning error mainly depends by the underwater acoustic localization system the added value of the developed markerless hybrid approach lies in its capability to interpolate two consecutive acoustic positioning data through VIO tracking techniques in a sufficiently accurate way to perform a consistent and smooth AR visualization.

In conclusions, in cases in which the acoustic localization system cannot be implemented because of environmental, technical or economic reasons, then the UWAR visualization can be experienced by divers by means the hybrid maker-based approach. However, although this is the most cost-effective solution in terms of hardware costs because it does not need the adoption of the acoustic localization system, the maker-based approach requires the area to be populated by artificial visual markers and to keep them clean since just a few days are sufficient to be covered with a thin layer of biofouling that makes them completely unintelligible. Otherwise, it is possible to perform UWAR according to the proposed markerless approach in which the adoption of hybrid tracking techniques allow to improve the update rate in order to provide a fluid AR visualization.

As future development of the proposed markerless approach, in the next experimentations, also SBL and LBL systems will be taken into consideration in order to verify if these solutions allow to improve the performance of the acoustic tracking in a shallow water context.

**Acknowledgments**

This work has been supported by the iMARECulture project that has received funding from the European Union's Horizon 2020 research and innovation programme under grant agreement No. 727153. The authors would like to thank the Marine Protected Area-Underwater Park of Baiae for the permission to conduct the experimentation, and the ISCR (Istituto Superiore per la Conservazione ed il Restauro) for the support, collaboration and data provided.